\documentclass[ALICE,manyauthors]{cernphprep}

\usepackage[comma,square,numbers,sort&compress]{natbib}
\usepackage{hyperref}
\usepackage{lineno}
\usepackage{booktabs}
\usepackage{tabularx}
\newcolumntype{C}{>{\centering\arraybackslash}X}
\usepackage{siunitx}
\usepackage{caption}

\begin{document}%

\newcommand {\red}[1]    {\textcolor{red}{#1}}
\newcommand {\green}[1]  {\textcolor{green}{#1}}
\newcommand {\blue}[1]   {\textcolor{blue}{#1}}

\newcommand{\deutsch}[1]{\foreignlanguage{german}{#1}}

\newcommand{\CZ}[1]{\blue{{\small \textbf{CZ:} #1}}}

\DeclareRobustCommand{\unit}[2][]{%
        \begingroup%
                \def\0{#1}%
                \expandafter%
        \endgroup%
        \ifx\0\@empty%
                \ensuremath{\mathrm{#2}}%
        \else%
                \ensuremath{#1\,\mathrm{#2}}%
        \fi%
        }
\DeclareRobustCommand{\unitfrac}[3][]{%
        \begingroup%
                \def\0{#1}%
                \expandafter%
        \endgroup%
        \ifx\0\@empty%
                \raisebox{0.98ex}{\ensuremath{\mathrm{\scriptstyle#2}}}%
                \nobreak\hspace{-0.15em}\ensuremath{/}\nobreak\hspace{-0.12em}%
                \raisebox{-0.58ex}{\ensuremath{\mathrm{\scriptstyle#3}}}%
        \else
                \ensuremath{#1}\,%
                \raisebox{0.98ex}{\ensuremath{\mathrm{\scriptstyle#2}}}%
                \nobreak\hspace{-0.15em}\ensuremath{/}\nobreak\hspace{-0.12em}%
                \raisebox{-0.58ex}{\ensuremath{\mathrm{\scriptstyle#3}}}%
        \fi%
}

%
%
\newcommand{\ie}{i.\,e.\;}
\newcommand{\eg}{e.\,g.\;}

\newcommand{\run}[1]{\textsc{Run\,#1}}

\newcommand{\pp}{\ensuremath{\mathrm {p\kern-0.05em p}}\xspace}
\newcommand{\PbPb}{\ensuremath{\mbox{Pb--Pb}}\xspace}
\newcommand{\pPb}{\ensuremath{\mbox{p--Pb}}\xspace}

%
%

\newcommand{\dEdx}{\ensuremath{\mathrm{d}E/\mathrm{d}x}\xspace}
\newcommand{\dNdeta}{\ensuremath{\mathrm{d}N_{\rm ch}/\mathrm{d}\eta}\xspace}
\newcommand{\lambdac}{\ensuremath{\Lambda_{\rm c}^{+}}}
\newcommand{\lambdab}{\ensuremath{\Lambda_{\rm b}^{0}}}
\newcommand{\xicz}{\ensuremath{\Xi_{\rm c}^{0}}}
\newcommand{\xiczp}{\ensuremath{\Xi_{\rm c}^{0,+}}}
\newcommand{\xicp}{\ensuremath{\Xi_{\rm c}^{+}}}
\newcommand{\xib}{\ensuremath{\Xi_{\rm b}}}
\newcommand{\pt}{\ensuremath{p_{\mathrm{T}}}}
\newcommand{\Dzero}{\ensuremath{\mathrm {D^0}}\xspace}
\newcommand{\Dzerobar}{\ensuremath{\mathrm{\overline{D}^0}}\xspace}
\newcommand{\xiczsemilep}{\ensuremath{\xicz\ \rightarrow {\rm e}^+\Xi^-\nu_{\rm e}}}
\newcommand{\exipair}{\ensuremath{{\rm e}^+\Xi^-}}

%
%
\newlength{\smallerpicsize}
\setlength{\smallerpicsize}{70mm}
\newlength{\smallpicsize}
\setlength{\smallpicsize}{90mm}
\newlength{\mediumpicsize}
\setlength{\mediumpicsize}{120mm}
\newlength{\largepicsize}
\setlength{\largepicsize}{150mm}

\newcommand{\PICX}[5]{
   \begin{figure}[!hbt]
      \begin{center}
         \vspace{3ex}
         \includegraphics[width=#3]{#1}
         \caption[#4]{\label{#2} #5}        
      \end{center}  
   \end{figure}
}

\newcommand{\PICH}[5]{
   \begin{figure}[H]
      \begin{center}
         \vspace{3ex}
         \includegraphics[width=#3]{#1}
         \caption[#4]{\label{#2} #5}        
      \end{center}  
   \end{figure}
}

%
%
%
\newcommand{\figs}{Figs.\xspace}
\newcommand{\Figs}{Figures\xspace}
\newcommand{\eqn}{equation\xspace}
\newcommand{\Eqn}{Equation\xspace}
\newcommand{\figref}[1]{Fig.~\ref{#1}}
\newcommand{\Figref}[1]{Figure~\ref{#1}}
\newcommand{\tabref}[1]{Tab.~\ref{#1}}
\newcommand{\Tabref}[1]{Table~\ref{#1}}
\newcommand{\appref}[1]{appendix~\ref{#1}}
\newcommand{\Appref}[1]{Appendix~\ref{#1}}
\newcommand{\secs}{Secs.\xspace}
\newcommand{\Secs}{Sections\xspace}
\newcommand{\secref}[1]{Sec.~\ref{#1}}
\newcommand{\Secref}[1]{Section~\ref{#1}}
\newcommand{\chaps}{Chaps.\xspace}
\newcommand{\Chaps}{Chapters\xspace}
\newcommand{\chapref}[1]{Chap.~\ref{#1}}
\newcommand{\Chapref}[1]{Chapter~\ref{#1}}
\newcommand{\lstref}[1]{Listing~\ref{#1}}
\newcommand{\Lstref}[1]{Listing~\ref{#1}}
%
%
\newcommand{\otoprule}{\midrule[\heavyrulewidth]}
\topfigrule

\begin{titlepage}
\PHyear{2017}
\PHnumber{332}      
\PHdate{11 December}  
%

\title{First measurement of \xicz\ production in pp collisions at $\mathbf{\sqrt{s}}$ = 7 TeV }
\ShortTitle{$\Xi_{\rm c}^0$ production in pp collisions at $\sqrt{s}$ = 7 TeV }   

\Collaboration{ALICE Collaboration\thanks{See Appendix~\ref{app:collab} for the list of collaboration members}}
\ShortAuthor{ALICE Collaboration} 

\begin{abstract}
The production of the charm-strange baryon \xicz\  is  measured for the first time at the LHC via its  semileptonic decay into e$^+\Xi^-\nu_{\rm e}$ in \pp collisions at $\sqrt{s}=7$ TeV with the ALICE detector.  The transverse momentum (\pt) differential cross section multiplied by the branching ratio is presented in the interval 1 $<$ \pt\ $<$ 8~GeV/$c$ at mid-rapidity, $|y|$ $<$ 0.5. The transverse momentum dependence of the  \xicz\ baryon production relative to the \Dzero meson production is compared to predictions of event generators with various tunes of the hadronisation mechanism, which are found to underestimate the measured cross-section ratio.
\end{abstract} 
\end{titlepage}
\setcounter{page}{2}

%
%

Quantum Chromodynamics (QCD)  as the theory of the strong interaction  has been a cornerstone of the Standard Model for several decades. It has been tested  through measurements in e$^+$e$^-$, pp, $\mathrm{p}\overline{\mathrm{p}}$ and ep collisions at momentum-transfer scales where perturbative techniques are applicable~\cite{qcd_book}.  
In particular, measurements of charm hadrons have provided important tests of the theory because perturbative techniques are applicable down to low transverse momentum (\pt) thanks to the large mass of the charm quark compared to the QCD scale parameter ($\Lambda_{\rm QCD}\sim200$ MeV).  
The production cross sections of charm hadrons can be calculated using the factorisation approach as a convolution of three factors~\cite{Collins:1985gm}: the parton distribution functions of the incoming protons, the hard-scattering cross section at partonic level and the fragmentation functions of charm quarks into charm hadrons. 
There are several state-of-the-art calculations adopting different factorisation schemes. The collinear factorisation scheme is used by calculations at next-to-leading order  in $\alpha_s$, such as the general-mass variable flavour number scheme ({\sc gm-vfns})~\cite{Kniehl:2004fy,Kniehl:2005mk,Kniehl:2012ti} and the fixed order with next-to-leading-log resummation ({\sc fonll})~\cite{Cacciari:1998it,Cacciari:2012ny} approaches, while the $k_{\rm T}$ factorisation scheme is employed at leading order in  Refs.~\cite{Catani:1990eg,Luszczak:2008je,Maciula:2013wg}. 
However, some of these calculations do not  provide predictions for heavy-baryon production due to the lack of knowledge about the fragmentation function of charm quarks into baryonic states. 
Measurements of the production of charm  baryons, such as \lambdac\ and \xicz, are essential to develop and test models of the hadronisation process. 

While a variety of new charm-baryon resonances, such as $\Omega_{\rm c}^0$~\cite{Aaij:2017nav}, $\Xi_{\rm cc}^{++}$~\cite{Aaij:2017ueg}, have recently been found,  charm-hadron cross-section measurements  at the Large Hadron Collider (LHC) are mainly limited to mesons~\cite{Abelev:2012xe,Abelev:2012qh,Abelev:2012pi,Aaij:2013mga,Abelev:2014gla,Aaij:2015bpa,Aad:2015zix,Aaij:2016jht,Acharya:2017jgo}, apart from a few  measurements of the \lambdac\ cross section in \pp and \pPb collisions~\cite{Aaij:2013mga,LHCb-CONF-2017-005}. 
In the case of \xicz, the existing measurements are currently limited to e$^+$e$^-$ collisions~\cite{
Albrecht:1990zk,Albrecht:1992jt,Alexander:1994hp,Albrecht:1994hr,Aubert:2005cu}.
 New measurements of charm-baryon production are therefore needed to provide further insights into the hadronisation processes in \pp collisions. 
For example, interactions at the partonic level among the produced quarks and gluons, such as colour reconnection, could be stronger in \pp collisions than  in e$^+$e$^-$ collisions, resulting in an enhanced production of  baryons relative to mesons \cite{Christiansen:2015yqa}. 
The measurements of charm-baryon production in \pp collisions also serve as a reference for heavy-ion collisions, where a modification of the baryon-to-meson ratio is expected   if a substantial fraction of charm quarks hadronises via recombination with other quarks from the deconfined medium created in the collision~\cite{Sorensen:2005sm,Lee:2007wr,MartinezGarcia:2007hf,Oh:2009zj,Ghosh:2014oia}. 
Measurements of charm-strange baryons, \eg  \xicz, could also provide additional input to better understand the hadronisation mechanism of strange quarks in \pp collisions because of their valence quark composition.   


In this paper, we report the first measurement  of the \pt-differential production cross section of \xicz\  multiplied by the branching ratio (BR) into the semileptonic decay mode, \xiczsemilep, and its ratio to the measured production cross section of \Dzero mesons~\cite{Acharya:2017jgo} as a function of \pt,  up to 8 GeV/$c$. 
The absolute branching ratio of this \xicz\ decay is currently unknown~\cite{Agashe:2016kda}.   
Using a data sample of \pp collisions at $\sqrt{s} = 7$ TeV recorded with the ALICE detector in 2010, the measurement is performed by analyzing \exipair\ pairs formed by combining positrons and $\Xi^-$ baryons reconstructed with the detectors of the ALICE central barrel, covering the pseudorapidity interval $|\eta| < 0.9$. 
The missing momentum of the neutrino is corrected using unfolding techniques. 
Charge conjugate modes are implied everywhere, unless otherwise stated. 
Only the sub-detectors relevant for this data analysis are described below. A more complete and detailed description of the ALICE detector and its performance can be found in Refs.~\cite{Aamodt:2008zz, Abelev:2014ffa}. 

The detectors used in this analysis include the Inner Tracking System (ITS), the Time Projection Chamber (TPC) and the Time-Of-Flight detector (TOF). These detectors are located in a large solenoid magnet producing a magnetic field of 0.5 T parallel to the LHC beam axis.  
The ITS consists of six cylindrical layers of silicon detectors, placed at radial distances ranging from 3.9 cm to 43 cm from the nominal beam axis and covering the full azimuth.
The two innermost layers consist of  Silicon Pixel Detectors (SPD), the two intermediate layers of  Silicon Drift Detectors (SDD) and the two outermost layers of Silicon Strip Detectors (SSD).   
The total material budget of the ITS is on average 7.7\% of a radiation length, for particles with $\eta=0$~\cite{Aamodt:2010aa}. The ITS spatial resolution enables the measurement of the distance of closest approach ($d_0$) of tracks to the primary vertex with a resolution better than \SI{75}{\micro\meter} in the transverse plane for \pt\ $>$ 1~GeV/$c$ in \pp collisions~\cite{ALICE:2011aa}. 
The TPC is a cylindrical gaseous detector with a volume of about 90 m$^3$. 
The TPC provides track reconstruction with up to 159 space points at radial distances from the beam axis  ranging between 85 cm and 247 cm, within the full azimuth.  
The TPC cluster-position resolution is about \SI{500}{\micro\meter} along the beam direction and in the transverse direction for tracks with $\eta=0$~\cite{Alme:2010ke}. 
The TPC also provides particle identification capabilities via the measurement of the specific ionisation energy loss, \dEdx, with a resolution of approximately 5.2\% in \pp collisions~\cite{Abelev:2014ffa}. The TOF detector consists of multi-gap resistive plate chambers placed at a radial distance of 3.7 m from the beam axis and also covers the full azimuth. 
The TOF detector, with a timing resolution of about 80~ps,   measures the time-of-flight of particles relative to the time of the collision, which is determined by
 the arrival time of the particles at the TOF detector and by the T0 detector, an array of Cherenkov counters placed at +370 cm and $-$70~cm from the nominal interaction point along the beam axis~\cite{Adam:2016ilk}.

The analysed data sample consists of \pp collisions at $\sqrt{s}$ = 7 TeV recorded during the  2010 LHC data taking  period with a minimum bias trigger that requires at least one hit in either the SPD or the V0 detectors. The two layers of the SPD detector cover $|\eta|<2.0$. The two V0 detectors, each comprising 32 scintillator tiles, are installed on both sides of the interaction point and cover $-3.7 < \eta < -1.7$ and $2.8 < \eta < 5.1$.  The trigger condition captures 87\% of the \pp inelastic cross section~\cite{Abelev:2012sea}. The collision vertex is reconstructed with an efficiency of 88\% and only events with a reconstructed vertex within 10~cm from the nominal interaction point along the beam direction are used in this analysis. Pile-up events are identified by searching for a second interaction vertex, reconstructed with at least three SPD tracklets (that are two-point track segments connecting hits in the two SPD layers) pointing to a common vertex, which is separated from the first vertex by at least 8 mm.  
After the selections, the analysed sample corresponds to an integrated luminosity $L_{\rm int}$ = 5.9 $\pm$ 0.2 nb$^{-1}$. 
 

The \xicz\ candidates are defined from \exipair\ pairs by combining a track originating from the primary vertex (denoted by ``electron track'' in the following) and a reconstructed $\Xi^-$ baryon. 
Electron tracks satisfying $|\eta|<0.8$ and $\pt>0.5$ GeV/$c$ are required to have at least 100 associated clusters in the TPC (out of which at least 80 are used for the calculation of the \dEdx signal), a $\chi^2$ normalised to the number of TPC clusters smaller than 4 and at least 4 hits in the ITS. It is also required that the electron track has associated hits in the two innermost layers of the ITS, in order to reject electrons from photon conversions occurring in the detector material outside the innermost SPD layer~\cite{Abelev:2012xe}. Electrons are identified using the \dEdx measurement in the TPC and the time-of-flight measurement of the TOF detector. 
In both cases, the selection is applied on the $n_\sigma^{\mathrm{TPC}}$ and $n_\sigma^{\mathrm{TOF}}$ variables defined as the difference between the measured \dEdx or time-of-flight values and the one expected for electrons, divided by the corresponding detector resolution.
The following selection criteria are applied: $|n_\sigma^{\mathrm{TOF}}| < 3$ and $-3.9+1.2\pt-0.094\pt^2<n_\sigma^{\mathrm{TPC}} (\pt) < 3$.  The \pt-dependent lower limit on $n_\sigma^{\mathrm{TPC}}$ was optimised to reject hadrons. Thus, an electron purity of 98\% is achieved over the whole \pt\  range. 

The background from ``photonic'' electrons (originating from Dalitz decays of neutral mesons and photon conversions in the detector material) remaining in the electron sample are identified using a technique based on the invariant mass of e$^+$e$^-$ pairs~\cite{Adam:2015qda}. The electron tracks are paired with opposite-sign tracks from the same event passing  loose selection criteria  ($|n_\sigma^{\mathrm{TPC}}| <5$ without TOF requirement) and are identified as photonic electrons if there is at least one pair with an invariant mass smaller than 50 MeV/$c^2$. Setting such loose  electron identification criteria is meant to increase the efficiency of finding the partners. This improves the signal-to-background ratio for \xicz\ by about 50\%, while the fraction of the signal lost due to misidentifications is less than 2 \%. 

\begin{figure}[htbp]
\begin{center}
\includegraphics[width=7.9cm]{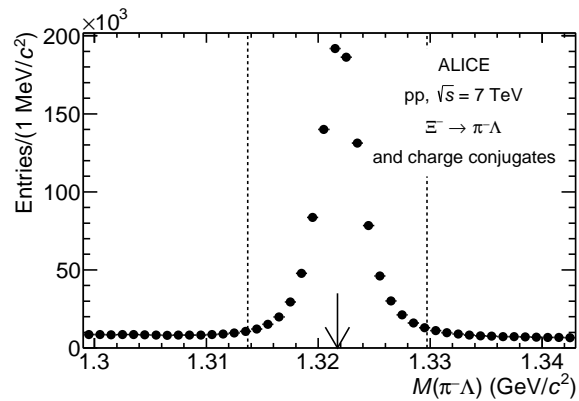}
\caption{Invariant-mass distribution of  $\Xi^- \rightarrow \pi^-\Lambda$ (and charge conjugate) candidates integrated over \pt. The arrow indicates the world average $\Xi^-$ mass from Ref.~\cite{Agashe:2016kda} and the dashed lines indicate the selected interval for the $\Xi^-$ candidates. }
\label{fig:xi_raw_incpt}
\end{center}
\end{figure}

The $\Xi^-$ baryons are reconstructed from the decay chain $\Xi^-\rightarrow \pi^-\Lambda$, followed by $\Lambda\rightarrow {\rm p}\pi^-$. Tracks used to define $\Xi^-$ candidates are required to have at least 80  clusters in the TPC and a \dEdx signal in the TPC  consistent with the expected values for protons (pions) within 4$\sigma$. The $\Xi^-$ and $\Lambda$ baryons have long lifetimes ($c\tau$ of about 4.91 cm and 7.89 cm, respectively~\cite{Agashe:2016kda}), and thus they can be identified using their characteristic cascade-like or V-shaped decay topologies \cite{Aamodt:2011zza,Abelev:2012jp,Abelev:2014qqa}. Pions originating directly from $\Xi^-$ decays are selected by requiring $d_0 > 0.02$ cm; protons and pions originating from $\Lambda$ decays are required to have $d_0 > 0.07$ cm.  The $d_0$ of the $\Lambda$ trajectory to the primary vertex is required to be larger than 0.03 cm, while its cosine of the pointing angle, which is the angle between the reconstructed $\Lambda$ momentum and the line connecting the $\Lambda$ and $\Xi^-$ decay vertices, is required to be larger than 0.98. The distances of the $\Xi^-$ and $\Lambda$ decay vertices from the beam line are required to be larger than 0.4 and 2.7 cm, respectively. 
These selection criteria are tuned to reduce the background, while keeping a high efficiency for the signal. 
Figure \ref{fig:xi_raw_incpt} shows the $\Xi^-$ peak in the $\pi^-\Lambda$ invariant-mass distribution integrated over \pt. 
Only $\Xi^-$ candidates with invariant masses within 8 MeV/$c^2$ from the $\Xi^-$ mass ($1321.71\pm0.07$~MeV/$c^2$~\cite{Agashe:2016kda}) indicated by an arrow in Figure \ref{fig:xi_raw_incpt} are kept for further analysis. In this interval, the signal-to-background ratio is about 8. 

The \exipair\ pairs are formed from selected positrons and $\Xi^-$ candidates. Only pairs with an opening angle smaller than 90 degrees are used for the analysis. 
The background in the \exipair\ pair distribution 
is estimated by exploiting the fact that \xicz\ baryons decay into  ${\rm e}^+\Xi^-\nu_{\rm e}$ (right-sign, RS), but not into ${\rm e}^-\Xi^-\overline{\nu}_{\rm e}$ (wrong-sign, WS), while most of the background sources contribute equally to RS and WS pairs.  
The yield of WS pairs is therefore used to estimate the background and is subtracted from the yield of RS pairs to obtain the \xicz\ raw yield. 
The procedure is verified with {\sc pythia} 6.4.21~\cite{Sjostrand:2006za} simulations using the  Perugia-0 tune~\cite{Skands:2009zm} and the {\sc geant3} transport code~\cite{Brun:1994aa}, including a  realistic description of the detector response and alignment during the data taking period. 
A similar procedure was adopted by the ARGUS and CLEO collaborations studying e$^+$e$^-$ collisions~\cite{Albrecht:1992jt,Alexander:1994hp}. 

\begin{figure}[htbp]
\begin{center}
\includegraphics[width=7.9cm]{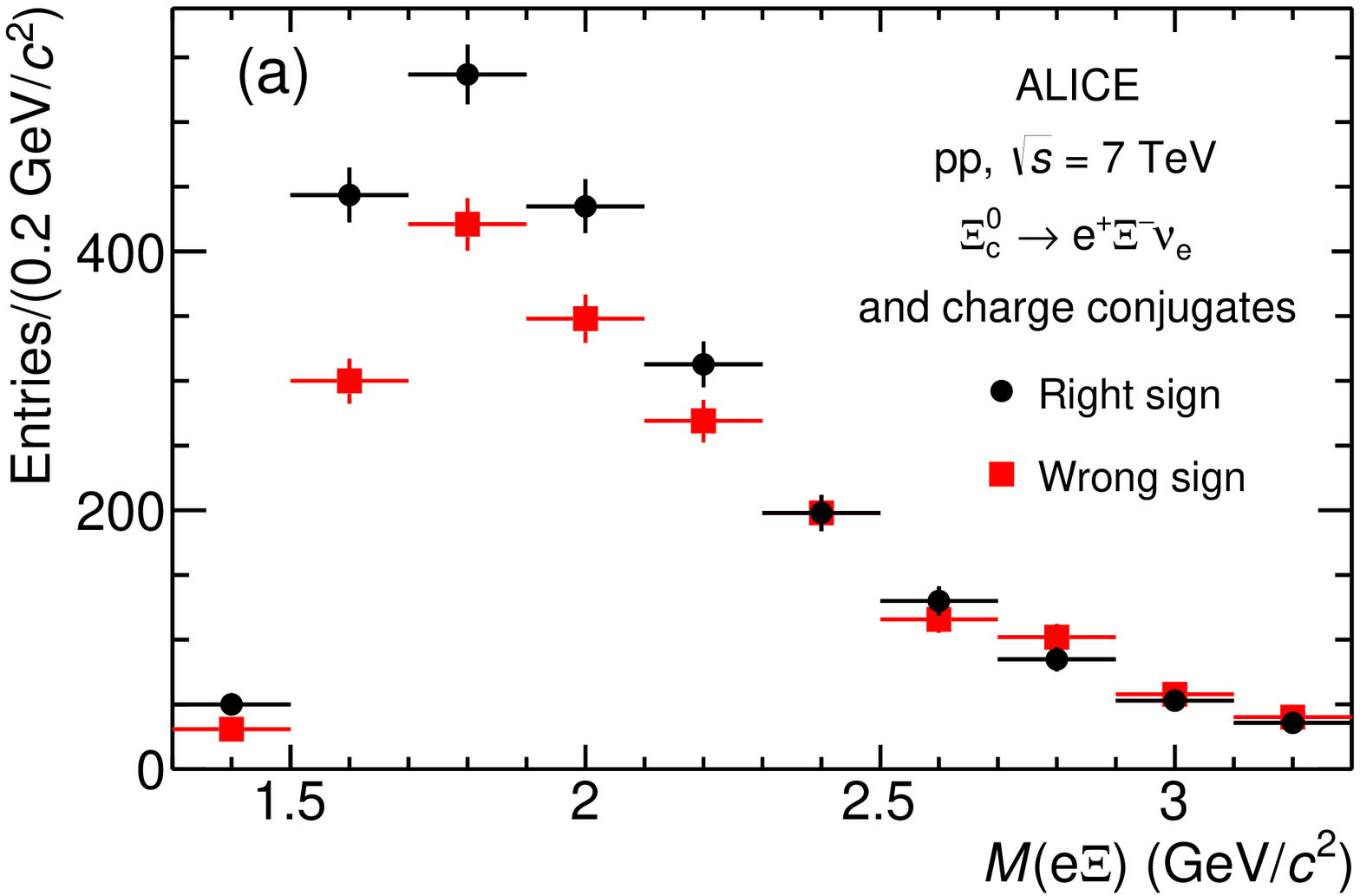}
\includegraphics[width=7.9cm]{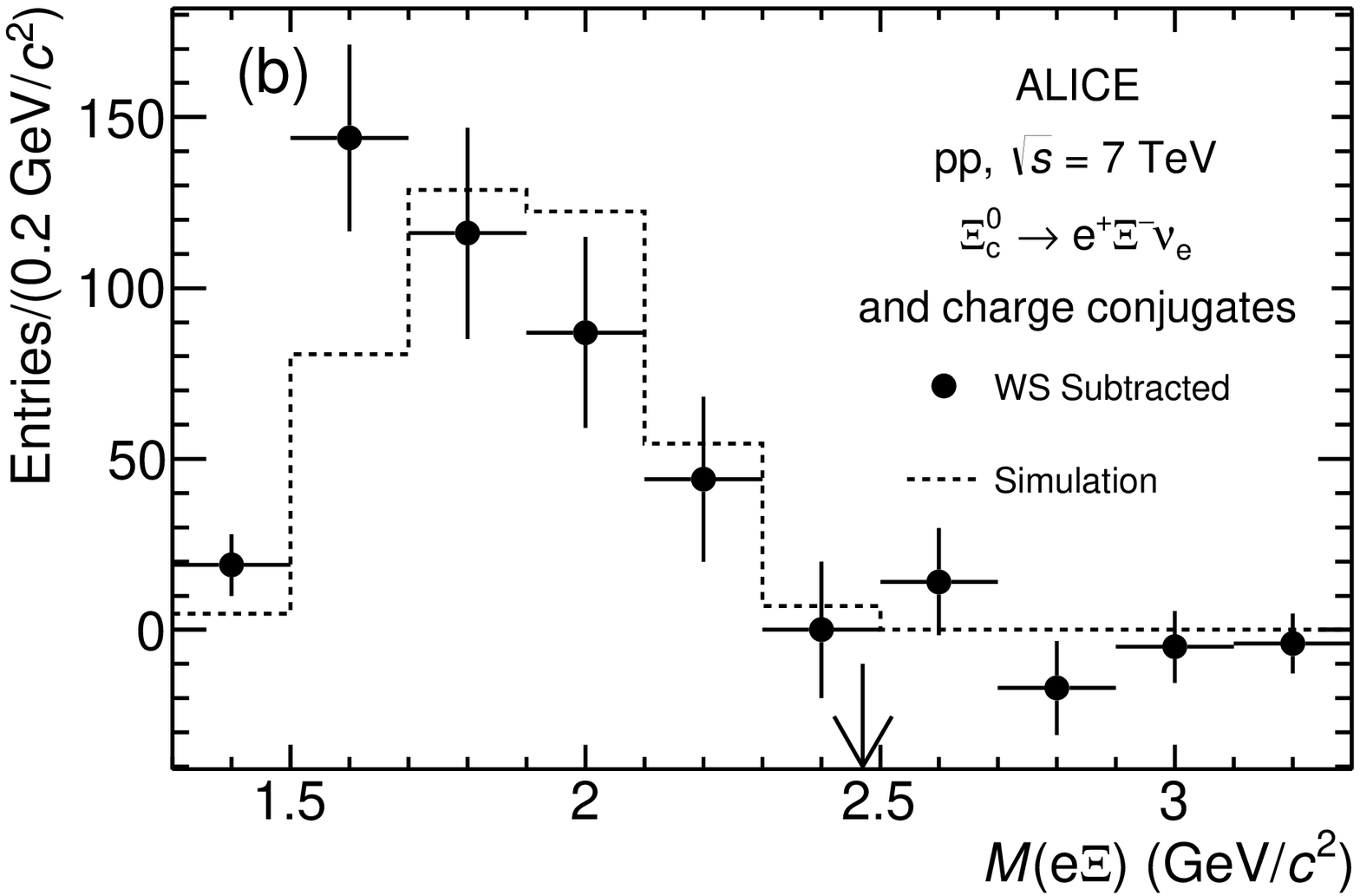}
\caption{(a) Invariant-mass distributions of right-sign and wrong-sign (and charge conjugate) pairs integrated over the whole \pt\ interval. (b)  Invariant-mass distribution of \xicz\ candidates obtained by subtracting the wrong-sign pair yield from the right-sign one compared with the signal distribution from the simulation, which is normalised to the measured RS$-$WS yield.  The arrow indicates the \xicz\ mass~\cite{Agashe:2016kda}.}
\label{fig:elexi_raw_incpt}
\end{center}
\end{figure}

Figure \ref{fig:elexi_raw_incpt}(a) shows the invariant-mass distributions of RS and WS pairs, integrated over the whole \pt\ interval.    
The invariant-mass distribution of \xicz\ candidates  obtained by subtracting the WS pair yield from the RS one is shown in Figure ~\ref{fig:elexi_raw_incpt}(b) together with the signal distribution from the simulation, which is normalised to the measured RS$-$WS yield. 
The shapes of the two distributions are found to be consistent with each other. 
Due to the missing momentum of the neutrino, the invariant-mass distribution of the \exipair\  pair does not peak at the \xicz\ mass ($2470.85^{+0.28}_{-0.40}$~MeV/$c^2$~\cite{Agashe:2016kda}) indicated by an arrow in Figure~\ref{fig:elexi_raw_incpt}(b).  
The invariant mass of \exipair\ pairs from \xicz\ decays is bounded by the \xicz\ mass due to the missing momentum of the neutrino. Thus only \exipair\ pairs satisfying $m_{{\rm e}\Xi}$ $<$ 2.5 GeV/$c^2$ are selected for further analysis. 

In order to obtain the \pt-differential production cross section of \xicz\ baryons, the background-subtracted (WS-subtracted) yield needs to be corrected for: the signal loss due to misidentification of photonic electrons, the \xib\ contribution in the WS pairs, the missing neutrino momentum, the detector acceptance and the track-reconstruction and the candidate-selection efficiencies. 
No correction is applied for possible differences in the acceptance of RS and WS pairs, which are found to be negligible for the current analysis based on a study with the mixed-event technique (\ie by pairing electrons and $\Xi^-$ from different events).

The first correction accounts for the signal loss caused by the misidentification of photonic electrons. 
The misidentification occurs when electrons from \xicz\ decays accidentally have opposite-sign partners giving rise to a very small invariant mass of the e$^+$e$^-$ pair. 
The misidentification probability is estimated to be less than 2\% by applying the tagging algorithm  to  e$^+$e$^+$ and e$^-$e$^-$ pairs.  
The correction is applied as a function of the \pt\ of  the \exipair\ pair. 

The second correction accounts for the overestimation of the background caused by $\xib \rightarrow {\rm e}^-\Xi^- \overline{\nu}_{\rm e}X$ decays, which produce WS pairs.  
Since the branching ratio of \xib\ into ${\rm e}^-\Xi^- \overline{\nu}_{\rm e}X$ and the \xib\ cross section in \pp collisions at LHC energies have not been measured yet, two assumptions are made to estimate this contribution.  First, the shape of the transverse momentum distribution of the \xib\ baryon is assumed to be the same as that of \lambdab, which was measured for $\pt >$ 10 GeV/$c$ and $|y|<2$ by the CMS collaboration~\cite{Chatrchyan:2012xg}.  This measurement is extrapolated to \pt\ $=$ 0 using the Tsallis function,

\begin{equation}
C\pt\left[1+\frac{\sqrt{\pt^2+m^2}-m}{nT}\right] 
\end{equation}
 
whose parameters were also determined by the CMS collaboration by fitting the measured distribution.   The fit parameters are consistent with those determined by the LHCb collaboration for the measurement of  \lambdab\ down to \pt\ $=$ 0 at forward rapidity (2 $<$ $y$ $<$ 4.5)~\cite{Aaij:2015fea}.  The second hypothesis is made for the total yield of $\xib\rightarrow {\rm e}^-\Xi^-\overline{\nu}_{\rm e}X$, which is determined by using the measurements of
BR$({\rm b} \rightarrow \xib)\cdot{\rm BR}(\xib\rightarrow \Xi^- l^-\overline{\nu}X)$~\cite{Buskulic:1996sm} and 
 BR$({\rm b} \rightarrow \lambdab)\cdot{\rm BR}(\lambdab\rightarrow \Lambda l^-\overline{\nu}X)$~\cite{Barate:1997if} in e$^+$e$^-$ collisions and by assuming that the fraction of beauty quarks that hadronise into  \lambdab\ and \xib\ baryons are the same as those  in e$^+$e$^-$ collisions. 
This assumption is supported by B-meson measurements, which show that the yield of B$_{\rm s}^0$ mesons relative to non-strange B mesons is consistent in e$^+$e$^-$ and p$\overline{\rm p}$ collisions~\cite{Aaltonen:2008zd}.  
The \xib\ distribution obtained with these assumptions is further processed to take into account the detector acceptance, efficiency and the momentum carried by non-reconstructed decay particles. 
This is done with the {\sc pythia 6} simulation using {\sc geant3} for particle transport through the detector.   
The correction increases with \pt\ and reaches 2\% at the highest \pt\ interval. 

\begin{figure}[htbp]
\begin{center}
\includegraphics[width=11cm]{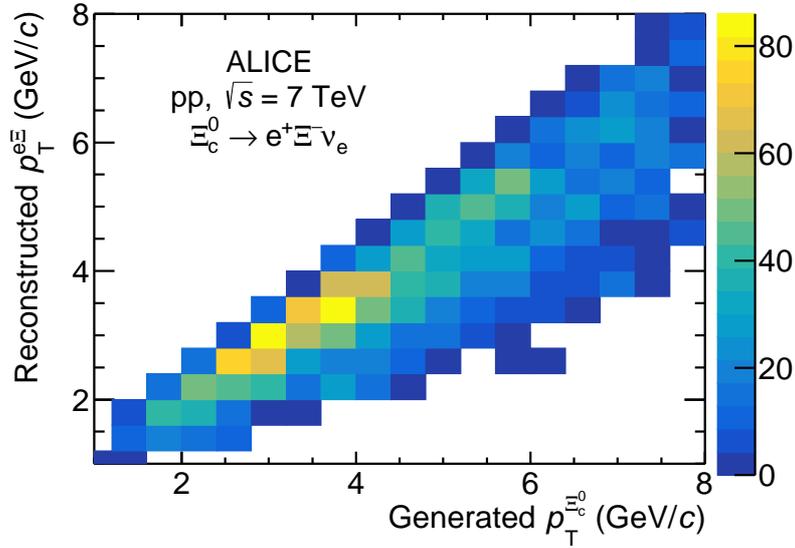}
\caption{Correlation between the generated \xicz-baryon \pt\ and the reconstructed \exipair\ pair \pt, obtained  from the  simulation based on {\sc pythia 6} described in the text. }
\label{fig:resp_xic}
\end{center}
\end{figure}

The transverse momentum distribution of \exipair\ pairs is corrected for the  missing momentum  of the neutrino using unfolding techniques. 
The response matrix to correct for the missing neutrino  momentum is generated based on the correlation between the \pt\ of the \xicz\ baryon and that of the reconstructed \exipair\ pair, which is obtained from the simulation described above and is shown in Figure \ref{fig:resp_xic}. 
The  response matrix includes both the decay kinematics and the instrumental effects, such as energy loss and bremsstrahlung in the detector material.   
The response matrix needs to be determined using a realistic \xicz-baryon \pt\ distribution.  However, the  distribution is not known {\it a priori}. Therefore, the response matrix is prepared in two steps. 
In the first step, the response matrix is obtained with the \pt\ distribution generated with {\sc pythia} 6. The resulting \xicz\ momentum distribution is used to produce the response matrix for the second iteration. 
The unfolding is performed with the RooUnfold ~\cite{Adye:2011gm} implementation of the Bayesian unfolding technique~\cite{D'Agostini:1994zf}, which is an iterative method based on Bayes' theorem. 
Convergence of the Bayesian method is achieved after three iterations.

The \pt-differential production cross section of \xicz\  baryons multiplied by the branching ratio into the considered semileptonic decay channel is calculated from the yields obtained by the unfolding  approach as follows:
\begin{eqnarray}
\label{eqn:xic_normalization}
\mathrm{BR}\cdot\frac{\mathrm{d}^2\sigma^{\xicz}}{\mathrm{d}\pt\mathrm{d}y} = \frac{ N_{\xicz}}{2\cdot\Delta \pt \Delta y\cdot (A\times\varepsilon)\cdot L_{\rm int}\cdot \mathrm{BR}_{\Xi^-}},
\end{eqnarray}
where $N_{\xicz}$ is the yield in a given $\pt$ interval with width $\Delta \pt$.  
The yield is divided by the integrated luminosity $L_{\rm int}$ of the analysed sample and by the product of the branching ratios of the decays $\Xi^-\rightarrow \pi^-\Lambda$ (99.887 $\pm$ 0.035\% ~\cite{Agashe:2016kda}) and $\Lambda\rightarrow {\rm p}\pi^-$ (63.9 $\pm$ 0.5\%~\cite{Agashe:2016kda}), which is indicated as BR$_{\Xi^-}$.  
The factor 1/2 is needed because the cross section is computed for the average of \xicz\ and $\overline{\Xi}_{\rm c}^0$, while the raw yield includes  both contributions. 
The factor $(A\times\varepsilon)$ is the product of the geometrical acceptance ($A$) and the reconstruction and selection efficiency ($\varepsilon$) for \xiczsemilep\ decays determined for \xicz\ generated in $|y|<0.8$. 
Finally, the yield is normalized to one unit of rapidity by dividing it by $\Delta y = 1.6$ under the assumption that the rapidity distribution of  \xicz\ is uniform in the range $|y|<0.8$.
 This assumption is verified with an accuracy of 1\% using {\sc pythia} 6. 
Note that the flatness of the rapidity distribution in $|y|<0.8$ is also relevant for the comparison to the \Dzero meson cross section, which was determined in $|y|<0.5$~\cite{Acharya:2017jgo}.

\begin{figure}[htbp]
\begin{center}
\includegraphics[width=11cm]{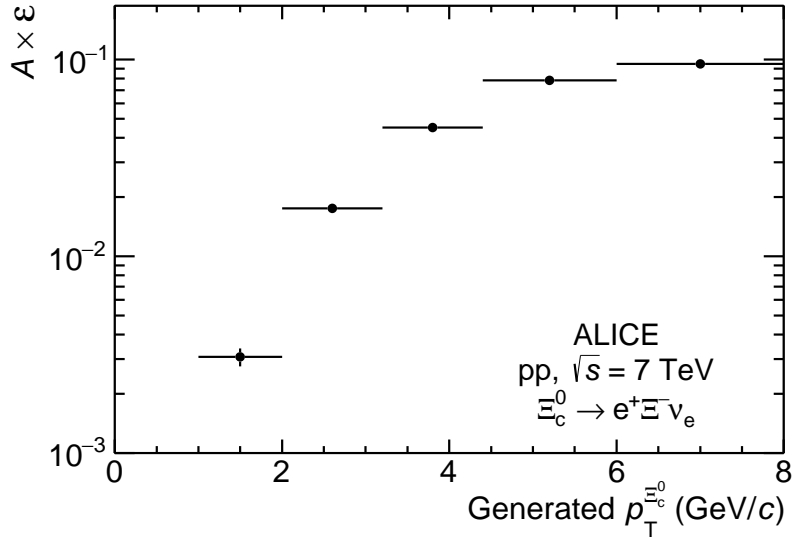}
\caption{Product of acceptance and efficiency ($A\times\varepsilon$) of \xicz\ baryons generated in $|y|<0.8$ decaying into ${\rm e}^+\Xi^-\nu_{\rm e}$  as a function of \pt, determined from  simulations {\sc pythia} 6 (see text). }
\label{fig:eff_xic}
\end{center}
\end{figure}

The acceptance and the efficiency are calculated from the simulations with an additional correction to take into account the fact that the elastic cross section of anti-protons is not accurate in {\sc geant3}~\cite{Abbas:2013rua}. 
The correction is calculated using the {\sc geant4} transport code~\cite{Agostinelli:2002hh}, which has a more accurate description of the cross section, and found to be less than 2\%. 
Since the acceptance and the efficiency depend on the \xicz-baryon \pt, the \xicz\  should be generated with a realistic momentum distribution. This was obtained via a two-step procedure similar to that used for the response matrix. 
Figure \ref{fig:eff_xic} shows the product of the geometrical acceptance and the reconstruction and selection efficiency $(A\times\varepsilon)$ of \xicz\ as a function of \pt.

The systematic uncertainty on the \xicz\ cross section has different contributions, which are the uncertainties on the raw yield (owing to the procedure of background estimation), on the $(A\times\varepsilon)$ factor (due to imperfections in the simulated samples), on the correction of the missing neutrino momentum (related  to the unfolding procedure) and on the normalisation.  
Table \ref{tab:syst_xic} summarises the estimated systematic uncertainties, reporting their values in all the \pt\ intervals. The total systematic uncertainty is determined by adding the individual contributions in quadrature in each \pt\ interval. 

The systematic uncertainty on the raw yield includes the uncertainties due to the WS subtraction procedure and  to the estimation of the \xib\ contribution.  
In the WS subtraction procedure described above, it was assumed that all the background sources contribute equally to RS and WS pairs.  
This is true as long as the background comprises uncorrelated pairs of electrons and $\Xi^-$. 
A systematic uncertainty of 4\% on the \xicz\ signal yield due to possible differences between RS and WS is estimated from simulations with the {\sc pythia} 6 event generator by checking the remaining contamination of background pairs in the RS yield after the subtraction of the WS pairs. 
The WS subtraction could also be affected by the amount of hadron contamination in the electron sample and the signal-to-background ratio of the \xicz\ signal. 
This effect is studied by repeating the analysis with different electron identification criteria. 
The results obtained with these modified criteria are found to be consistent with the ones from the default selections and therefore no systematic uncertainty is assigned.
The systematic uncertainty due to the \xib\ contribution to the WS pairs is estimated 
by varying the \xib\ momentum distribution within the quoted uncertainty of about 50\% on the cross section of \lambdab\ in \pp collisions  \cite{Chatrchyan:2012xg} and the quoted uncertainty of about 50\% on the ratio of the fragmentation fractions of beauty quarks into \lambdab\ and \xib\ in e$^+$e$^-$ collisions \cite{Buskulic:1996sm,Barate:1997if}.  
The effect on the final results is found to be about 1\% because the contribution from \xib\ is small. 
These systematic uncertainties add up to a total uncertainty of 5\% for the raw yield extraction. 

The systematic uncertainties arising from the reconstruction and selection efficiencies are estimated by repeating the analysis with different selection criteria for electrons, $\Xi^-$ and \exipair\ pairs  and by comparing the corrected yields. Due to the statistical limitations of the \xicz\ sample, the electron efficiencies are  studied via variations of the track-quality criteria and of the $n_\sigma$ values for the electron identification with TPC and TOF in the  $\lambdac\rightarrow {\rm e}^+\Lambda\nu_{\rm e}$ decays, which are analysed with the same procedure and have higher statistical significance. The RMS of the deviations of the corrected yields relative to the value obtained with the standard selection criteria, which amounts to 4\% and 3\%, is then assigned as a systematic uncertainty on the reconstruction and selection efficiency. Similarly, a systematic uncertainty of 1\% on both the $\Xi^-$ reconstruction and selection efficiency is estimated from the RMS deviation of the inclusive $\Xi^-$ corrected yield against variations of the criteria applied to select the $\Xi^-$ decay tracks and its cascade decay topology. 
In addition, a systematic uncertainty of 4\% on the $\Xi^-$ efficiency due to possible imperfections in the description of the detector material in the simulations~\cite{Abelev:2012jp} is considered and summed in quadrature with that estimated from the variation of the selection criteria.  
The uncertainties on the electron and $\Xi^-$ track-quality criteria are considered as correlated and combined linearly. 
The uncertainty on the \exipair\ pair selection efficiency is estimated by varying the selection criteria on the opening angle and the invariant mass of the pair and a systematic uncertainty of 3--27\% is assigned depending on \pt. 
Finally, a systematic uncertainty may also arise from an imperfect description of the acceptance of \exipair\ pairs in the simulation. It is estimated to be 11\% by comparing the azimuthal distributions of inclusive electrons and $\Xi^-$ baryons in the data and in the simulation. The uncertainty on the \exipair\ pair acceptance is summed in quadrature with that on the electron and $\Xi^-$ selection efficiencies, resulting in a systematic uncertainty on the $(A\times\varepsilon)$ correction factor ranging from 13\% to 30\% depending on \pt.

The systematic uncertainty on the missing neutrino momentum correction with the unfolding procedure is evaluated by varying the prior distribution to the Bayesian unfolding and by using different unfolding techniques, such as the $\chi^2$ minimisation method~\cite{Grosse-Oetringhaus:2011tgu,unfolding_chi2}  and the Singular Value Decomposition (SVD) method~\cite{Hocker:1995kb}. 
The RMS deviation of the results,  ranging between 4\% and 29\% depending on \pt, is assigned as a systematic uncertainty. 
A systematic uncertainty of 3\% is also assigned due to the imperfect knowledge of the \xicz-baryon \pt\ distributions used as input for the efficiency calculation and the unfolding procedure from the simulation. It is  estimated from the difference induced in the result by adding an additional step in the iterative procedure described above to obtain the input \pt\ distributions.
These systematic uncertainties add up to an uncertainty ranging between 6\% and 29\% depending on \pt. 

Finally, the results have a 3.5\% normalisation systematic uncertainty arising from the uncertainty in the determination of the minimum-bias trigger cross section in \pp collisions at $\sqrt{s}=7$ TeV \cite{Abelev:2012sea}. 

\begin{table}[t!]

\begin{center}
\begin{tabularx}{1.0\textwidth}{|C|CCCCC|}
\hline
 & \multicolumn{5}{ c| }{Relative systematic uncertainty (\%) in the measured \pt\ intervals (GeV/$c$)}\\
Source & 1--2 & 2--3.2 & 3.2--4.4 & 4.4--6 & 6--8 \\
\hline
Raw yield & 5  & 5 & 5 & 5 & 5\\
$(A\times\varepsilon)$ & 30 & 22 & 16 & 13 & 14 \\
$\pt^{\nu}$ & 29 & 8 & 6 & 7 & 10  \\
\hline
Normalisation & \multicolumn{5}{ c| }{3.5}\\
\hline
\end{tabularx}
\end{center}
\caption{Summary of systematic uncertainties on the \pt-differential cross section of $\xicz\ \rightarrow \rm {e}^+\Xi^-\nu_{\rm e}$ for 5 \pt\ intervals. The uncertainty on the missing neutrino momentum is denoted as $\pt^{\nu}$ in the table.}
\label{tab:syst_xic}
\end{table}%

\begin{figure}[ht]
\begin{center}
\includegraphics[width=11cm]{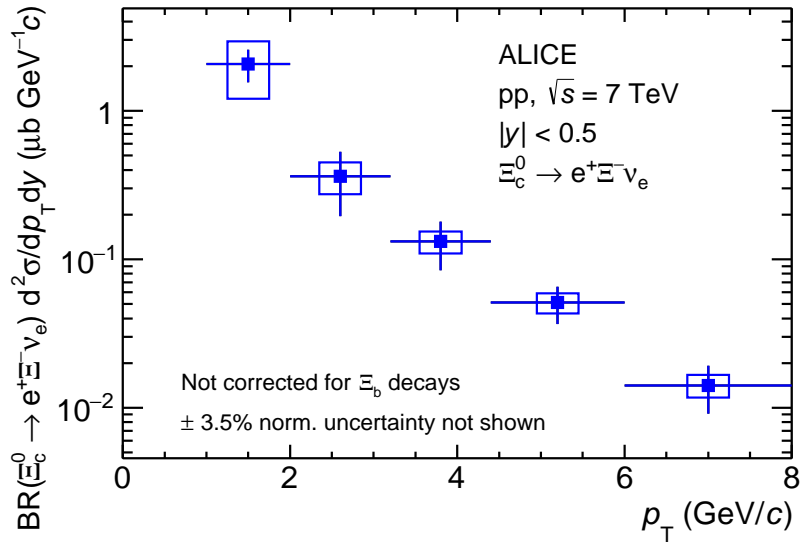}
\caption{Inclusive  \xicz-baryon \pt-differential production cross section multiplied by the branching ratio into ${\rm e}^+\Xi^-\nu_{\rm e}$, as a function of \pt\ for $|y|<0.5$, in \pp collisions at $\sqrt{s} =$ 7 TeV. The error bars and boxes represent the statistical and systematic uncertainties, respectively. The contribution from \xib\ decays is not subtracted. }
\label{fig:xic_spectra}
\end{center}
\end{figure}

 The \pt-differential cross section of \xicz\ baryons multiplied by the branching ratio into ${\rm e}^+\Xi^-\nu_{\rm e}$ is shown  in Figure \ref{fig:xic_spectra}  for the \pt\ interval $1 < \pt < 8$ GeV/$c$ at mid-rapidity, $|y|<0.5$. 
The error bars and boxes represent the statistical and systematic uncertainties, respectively. The feed down contribution from \xib, \eg $\xib^-\rightarrow \xicz\pi^-$ \cite{Aaij:2014lxa}, is not subtracted due to the lack of knowledge of the absolute branching ratios of $\xib\rightarrow\xicz+X$. 

\begin{figure}[h!]
\begin{center}
\includegraphics[width=9cm]{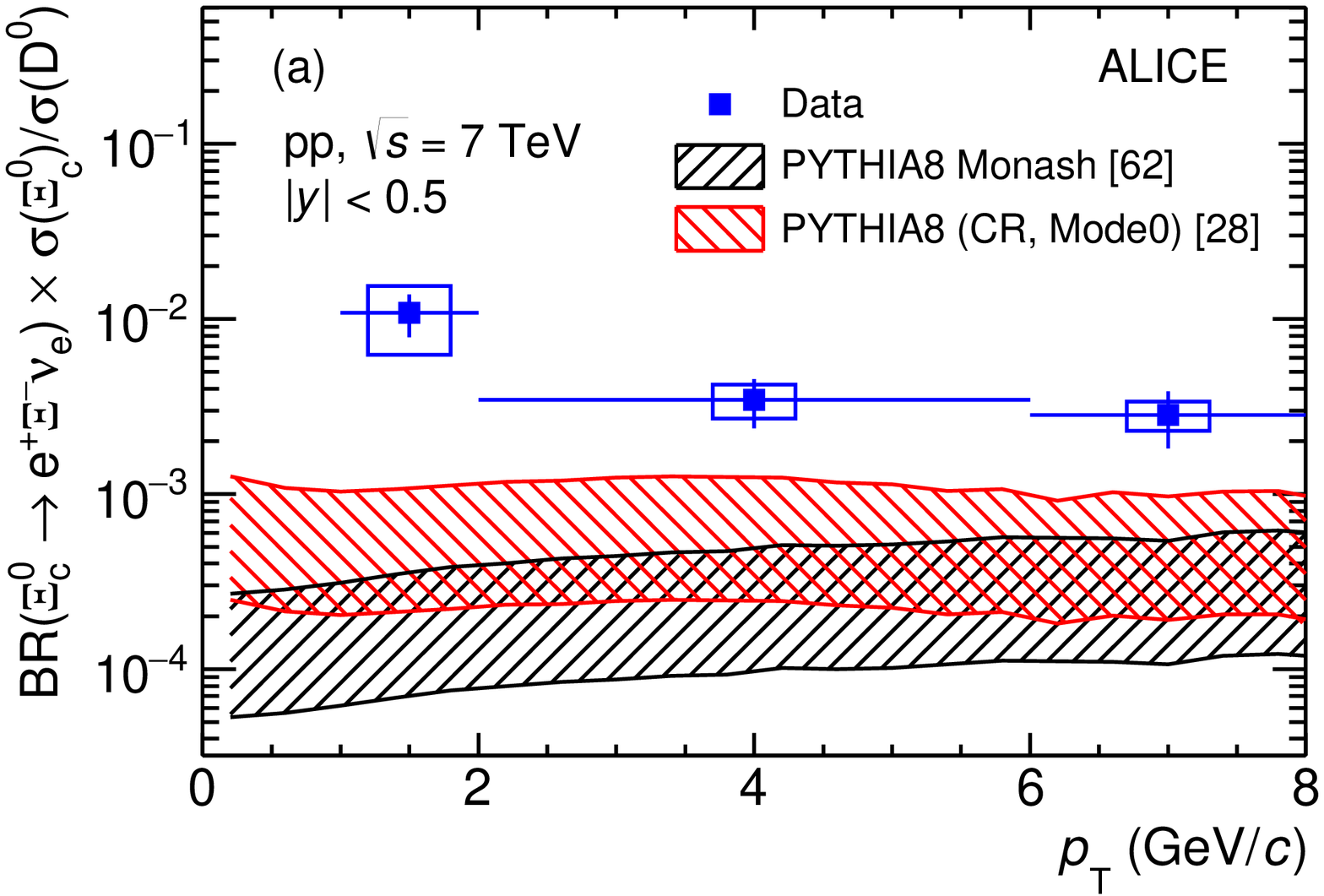}
\includegraphics[width=9cm]{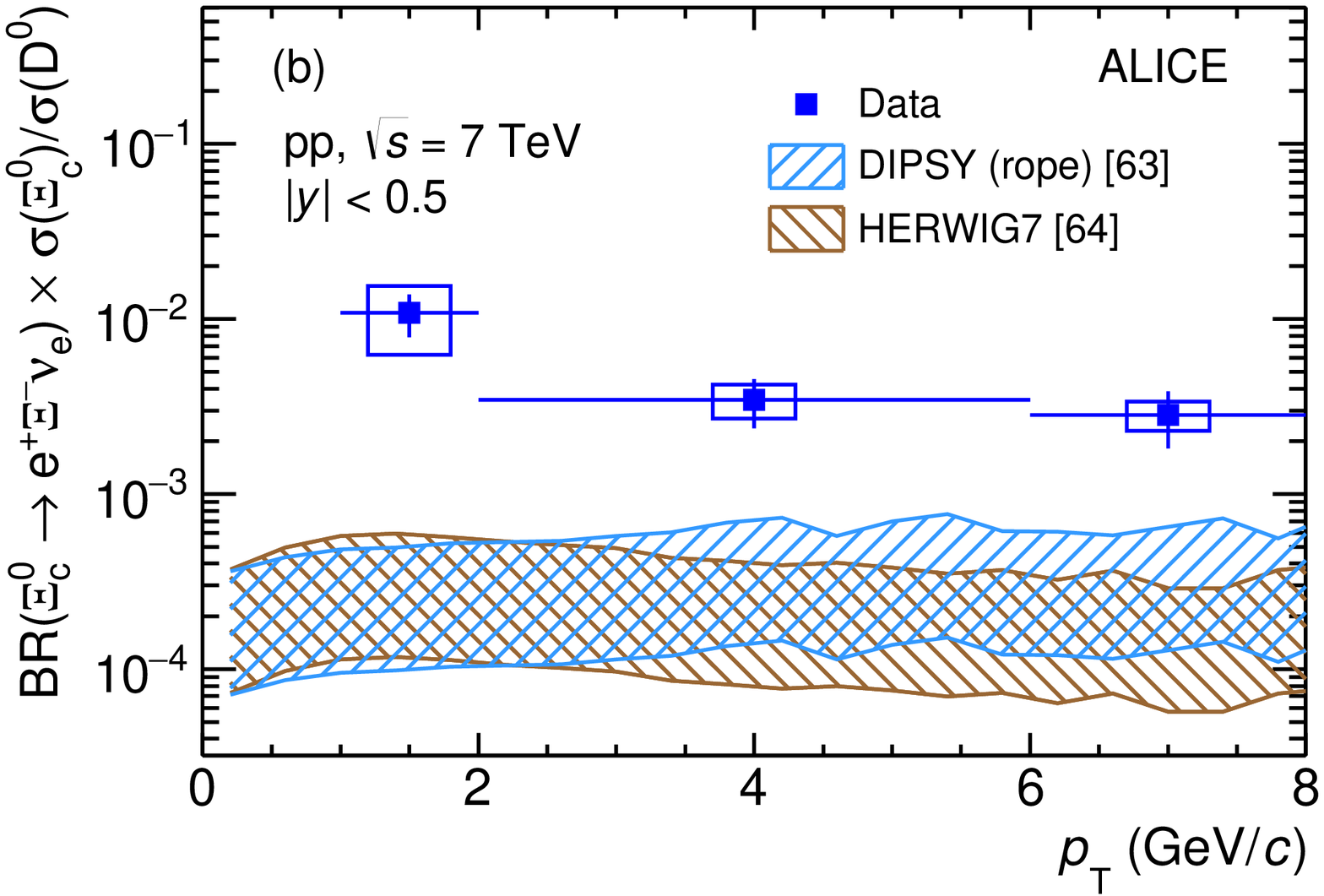}
\caption{
Ratio of the \pt-differential cross sections of \xicz\ baryons (multiplied by the branching ratio into ${\rm e}^+\Xi^-\nu_{\rm e}$) and \Dzero mesons~\cite{Acharya:2017jgo} as a function of \pt\  for $|y|<0.5$, in \pp collisions at $\sqrt{s} = $ 7 TeV. The error bars and boxes represent the statistical and systematic uncertainties, respectively. Predictions from theoretical models, (a) {\sc pythia 8} with different tunes \cite{Skands:2014pea,Christiansen:2015yqa} (b) {\sc dipsy}~\cite{Bierlich:2015rha} and {\sc herwig 7}~\cite{Bahr:2008pv},  are shown as shaded bands representing the range of the currently available theoretical predictions for the branching ratio of the considered \xicz\ decay mode. }
\label{fig:xic_d0_ratio}
\end{center}
\end{figure}

The ratio of the \pt-differential cross section of \xicz\ baryons to that of \Dzero mesons~\cite{Acharya:2017jgo}  is shown in Figure~\ref{fig:xic_d0_ratio}. 
The \pt\ intervals of the cross-section measurements are combined to have the same \pt\ bin boundaries for \xicz\ and \Dzero.  
The systematic uncertainty in a merged \pt\ interval is defined by propagating the yield extraction uncertainties of the \Dzero measurement as uncorrelated among \pt\ intervals and all the other uncertainties of the \Dzero and \xicz\ measurements as correlated.
The systematic uncertainty on the $\xicz/\Dzero$ ratio is calculated treating all the uncertainties on the \xicz\ and \Dzero cross sections as uncorrelated, except for the normalisation uncertainty that cancels out in the ratio. 
The ratio integrated in the transverse momentum interval $1<\pt<8$ GeV/$c$ is found to be $(7.0\pm1.5 (\mathrm{stat})\pm 2.6 (\mathrm{syst}))\times 10^{-3}$. 

In Figure \ref{fig:xic_d0_ratio}(a), the measured transverse momentum dependence of the $\xicz/\Dzero$ ratio is compared with  predictions from the {\sc pythia 8.211} event generator \cite{Sjostrand:2007gs,Sjostrand:2006za}. 
{\sc pythia 8} uses 2 $\rightarrow$ 2 processes followed by a leading-logarithmic \pt-ordered parton shower for the charm quark pair production and the hadronisation is treated with the Lund string model~\cite{Andersson:1983ia}.
The figure shows the results obtained with 
different tunes of hadronisation: the Monash 2013 tune~\cite{Skands:2014pea} and the Mode 0 tune from~\cite{Christiansen:2015yqa}. 
The latter  is based on a model for the hadronisation of multi-parton systems, which includes string formation beyond the leading-colour approximation and is implemented in {\sc pythia 8} with specific tuning of the colour reconnection parameters. As compared to the Monash 2013 tune, this model provides a better description of the measured baryon-to-meson ratios in the light-flavour sector.  
Two other tunes (Mode 2 and Mode 3)   provided in Ref.~\cite{Christiansen:2015yqa} give similar \xicz/\Dzero ratios as Mode 0.  
In Figure \ref{fig:xic_d0_ratio}(b), the measured ratio is also compared to other models implementing different hadronisation mechanisms: {\sc dipsy}~\cite{Bierlich:2015rha} with the rope hadronisation~\cite{Biro:1984cf} and {\sc herwig 7.0.4}~\cite{Bahr:2008pv} with the cluster hadronisation~\cite{Webber:1983if}. 
To compare the data with these models, theoretical calculations of the branching ratio, which range between 0.83\% and 4.2\%~\cite{PerezMarcial:1989yh,Singleton:1990ye,Cheng:1995fe}, are used. This range defines the width of the bands shown for the model calculations represented in Figure~\ref{fig:xic_d0_ratio}.  
Although the predictions of the Mode 0 tune of {\sc pythia} 8 are the closest to the data compared to the other models,  all  calculations underestimate the measured ratio significantly. 
Thus, this new measurement can provide an important constraint to the models of charm quark hadronisation in \pp collisions, once a measurement of the absolute branching ratio of the \xicz\ will become available.


In summary, we reported on the first LHC measurement of the inclusive \pt-differential production cross section of the charm-strange baryon \xicz\ multiplied by the branching ratio into ${\rm e}^+\Xi^-\nu_{\rm e}$ in \pp collisions at $\sqrt{s}$ = 7 TeV. 
The ratio of this measurement integrated over  $1<\pt<8$ GeV/$c$ to  the production cross section of the \Dzero\ meson integrated over the same \pt\ interval  was found to be  $(7.0\pm1.5 (\mathrm{stat})\pm 2.6 (\mathrm{syst}))\times 10^{-3}$. 
Several event generators with various models and tunes for the hadronisation mechanism underestimate the measured ratio. 
 
%
%

\newenvironment{acknowledgement}{\relax}{\relax}
\begin{acknowledgement}
\section*{Acknowledgements}

The ALICE Collaboration would like to thank all its engineers and technicians for their invaluable contributions to the construction of the experiment and the CERN accelerator teams for the outstanding performance of the LHC complex.
The ALICE Collaboration gratefully acknowledges the resources and support provided by all Grid centres and the Worldwide LHC Computing Grid (WLCG) collaboration.
The ALICE Collaboration acknowledges the following funding agencies for their support in building and running the ALICE detector:
A. I. Alikhanyan National Science Laboratory (Yerevan Physics Institute) Foundation (ANSL), State Committee of Science and World Federation of Scientists (WFS), Armenia;
Austrian Academy of Sciences and Nationalstiftung f\"{u}r Forschung, Technologie und Entwicklung, Austria;
Ministry of Communications and High Technologies, National Nuclear Research Center, Azerbaijan;
Conselho Nacional de Desenvolvimento Cient\'{\i}fico e Tecnol\'{o}gico (CNPq), Universidade Federal do Rio Grande do Sul (UFRGS), Financiadora de Estudos e Projetos (Finep) and Funda\c{c}\~{a}o de Amparo \`{a} Pesquisa do Estado de S\~{a}o Paulo (FAPESP), Brazil;
Ministry of Science \& Technology of China (MSTC), National Natural Science Foundation of China (NSFC) and Ministry of Education of China (MOEC) , China;
Ministry of Science, Education and Sport and Croatian Science Foundation, Croatia;
Ministry of Education, Youth and Sports of the Czech Republic, Czech Republic;
The Danish Council for Independent Research | Natural Sciences, the Carlsberg Foundation and Danish National Research Foundation (DNRF), Denmark;
Helsinki Institute of Physics (HIP), Finland;
Commissariat \`{a} l'Energie Atomique (CEA) and Institut National de Physique Nucl\'{e}aire et de Physique des Particules (IN2P3) and Centre National de la Recherche Scientifique (CNRS), France;
Bundesministerium f\"{u}r Bildung, Wissenschaft, Forschung und Technologie (BMBF) and GSI Helmholtzzentrum f\"{u}r Schwerionenforschung GmbH, Germany;
General Secretariat for Research and Technology, Ministry of Education, Research and Religions, Greece;
National Research, Development and Innovation Office, Hungary;
Department of Atomic Energy Government of India (DAE), Department of Science and Technology, Government of India (DST), University Grants Commission, Government of India (UGC) and Council of Scientific and Industrial Research (CSIR), India;
Indonesian Institute of Science, Indonesia;
Centro Fermi - Museo Storico della Fisica e Centro Studi e Ricerche Enrico Fermi and Istituto Nazionale di Fisica Nucleare (INFN), Italy;
Institute for Innovative Science and Technology , Nagasaki Institute of Applied Science (IIST), Japan Society for the Promotion of Science (JSPS) KAKENHI and Japanese Ministry of Education, Culture, Sports, Science and Technology (MEXT), Japan;
Consejo Nacional de Ciencia (CONACYT) y Tecnolog\'{i}a, through Fondo de Cooperaci\'{o}n Internacional en Ciencia y Tecnolog\'{i}a (FONCICYT) and Direcci\'{o}n General de Asuntos del Personal Academico (DGAPA), Mexico;
Nederlandse Organisatie voor Wetenschappelijk Onderzoek (NWO), Netherlands;
The Research Council of Norway, Norway;
Commission on Science and Technology for Sustainable Development in the South (COMSATS), Pakistan;
Pontificia Universidad Cat\'{o}lica del Per\'{u}, Peru;
Ministry of Science and Higher Education and National Science Centre, Poland;
Korea Institute of Science and Technology Information and National Research Foundation of Korea (NRF), Republic of Korea;
Ministry of Education and Scientific Research, Institute of Atomic Physics and Romanian National Agency for Science, Technology and Innovation, Romania;
Joint Institute for Nuclear Research (JINR), Ministry of Education and Science of the Russian Federation and National Research Centre Kurchatov Institute, Russia;
Ministry of Education, Science, Research and Sport of the Slovak Republic, Slovakia;
National Research Foundation of South Africa, South Africa;
Centro de Aplicaciones Tecnol\'{o}gicas y Desarrollo Nuclear (CEADEN), Cubaenerg\'{\i}a, Cuba and Centro de Investigaciones Energ\'{e}ticas, Medioambientales y Tecnol\'{o}gicas (CIEMAT), Spain;
Swedish Research Council (VR) and Knut \& Alice Wallenberg Foundation (KAW), Sweden;
European Organization for Nuclear Research, Switzerland;
National Science and Technology Development Agency (NSDTA), Suranaree University of Technology (SUT) and Office of the Higher Education Commission under NRU project of Thailand, Thailand;
Turkish Atomic Energy Agency (TAEK), Turkey;
National Academy of  Sciences of Ukraine, Ukraine;
Science and Technology Facilities Council (STFC), United Kingdom;
National Science Foundation of the United States of America (NSF) and United States Department of Energy, Office of Nuclear Physics (DOE NP), United States of America.
\end{acknowledgement}

\bibliographystyle{utphys}   
\bibliography{include/bibfile_sel}

\newpage
\appendix

\renewcommand{\thefigure}{\arabic{figure}}
\renewcommand{\thetable}{\arabic{table}}
\section{The ALICE Collaboration}
\label{app:collab}

\begingroup
\small
\begin{flushleft}
S.~Acharya\Irefn{org137}\And 
D.~Adamov\'{a}\Irefn{org94}\And 
J.~Adolfsson\Irefn{org34}\And 
M.M.~Aggarwal\Irefn{org99}\And 
G.~Aglieri Rinella\Irefn{org35}\And 
M.~Agnello\Irefn{org31}\And 
N.~Agrawal\Irefn{org48}\And 
Z.~Ahammed\Irefn{org137}\And 
S.U.~Ahn\Irefn{org79}\And 
S.~Aiola\Irefn{org141}
% ...
\And
\renewcommand\labelenumi{\textsuperscript{\theenumi}~}

\section*{Affiliation notes}
\renewcommand\theenumi{\roman{enumi}}
\begin{Authlist}
\item \Adef{org*}Deceased
\item \Adef{orgI}Dipartimento DET del Politecnico di Torino, Turin, Italy
\item \Adef{orgII}M.V. Lomonosov Moscow State University, D.V. Skobeltsyn Institute of Nuclear, Physics, Moscow, Russia
\item \Adef{orgIII}Department of Applied Physics, Aligarh Muslim University, Aligarh, India
\item \Adef{orgIV}Institute of Theoretical Physics, University of Wroclaw, Poland
\end{Authlist}

\section*{Collaboration Institutes}
\renewcommand\theenumi{\arabic{enumi}~}
\begin{Authlist}
\item \Idef{org1}A.I. Alikhanyan National Science Laboratory (Yerevan Physics Institute) Foundation, Yerevan, Armenia
\item \Idef{org2}Benem\'{e}rita Universidad Aut\'{o}noma de Puebla, Puebla, Mexico
\item \Idef{org3}Bogolyubov Institute for Theoretical Physics, Kiev, Ukraine
\item \Idef{org4}Bose Institute, Department of Physics  and Centre for Astroparticle Physics and Space Science (CAPSS), Kolkata, India
\item \Idef{org5}Budker Institute for Nuclear Physics, Novosibirsk, Russia
\item \Idef{org6}California Polytechnic State University, San Luis Obispo, California, United States
\item \Idef{org7}Central China Normal University, Wuhan, China
\item \Idef{org8}Centre de Calcul de l'IN2P3, Villeurbanne, Lyon, France
\item \Idef{org9}Centro de Aplicaciones Tecnol\'{o}gicas y Desarrollo Nuclear (CEADEN), Havana, Cuba
\item \Idef{org10}Centro de Investigaciones Energ\'{e}ticas Medioambientales y Tecnol\'{o}gicas (CIEMAT), Madrid, Spain
\item \Idef{org11}Centro de Investigaci\'{o}n y de Estudios Avanzados (CINVESTAV), Mexico City and M\'{e}rida, Mexico
\item \Idef{org12}Centro Fermi - Museo Storico della Fisica e Centro Studi e Ricerche ``Enrico Fermi', Rome, Italy
\item \Idef{org13}Chicago State University, Chicago, Illinois, United States
\item \Idef{org14}China Institute of Atomic Energy, Beijing, China
\item \Idef{org15}COMSATS Institute of Information Technology (CIIT), Islamabad, Pakistan
\item \Idef{org16}Departamento de F\'{\i}sica de Part\'{\i}culas and IGFAE, Universidad de Santiago de Compostela, Santiago de Compostela, Spain
\item \Idef{org17}Department of Physics, Aligarh Muslim University, Aligarh, India
\item \Idef{org18}Department of Physics, Ohio State University, Columbus, Ohio, United States
\item \Idef{org19}Department of Physics, Pusan National University, Pusan, Republic of Korea
\item \Idef{org20}Department of Physics, Sejong University, Seoul, Republic of Korea
\item \Idef{org21}Department of Physics, University of Oslo, Oslo, Norway
\item \Idef{org22}Department of Physics and Technology, University of Bergen, Bergen, Norway
\item \Idef{org23}Dipartimento di Fisica dell'Universit\`{a} 'La Sapienza' and Sezione INFN, Rome, Italy
\item \Idef{org24}Dipartimento di Fisica dell'Universit\`{a} and Sezione INFN, Cagliari, Italy
\item \Idef{org25}Dipartimento di Fisica dell'Universit\`{a} and Sezione INFN, Trieste, Italy
\item \Idef{org26}Dipartimento di Fisica dell'Universit\`{a} and Sezione INFN, Turin, Italy
\item \Idef{org27}Dipartimento di Fisica e Astronomia dell'Universit\`{a} and Sezione INFN, Bologna, Italy
\item \Idef{org28}Dipartimento di Fisica e Astronomia dell'Universit\`{a} and Sezione INFN, Catania, Italy
\item \Idef{org29}Dipartimento di Fisica e Astronomia dell'Universit\`{a} and Sezione INFN, Padova, Italy
\item \Idef{org30}Dipartimento di Fisica `E.R.~Caianiello' dell'Universit\`{a} and Gruppo Collegato INFN, Salerno, Italy
\item \Idef{org31}Dipartimento DISAT del Politecnico and Sezione INFN, Turin, Italy
\item \Idef{org32}Dipartimento di Scienze e Innovazione Tecnologica dell'Universit\`{a} del Piemonte Orientale and INFN Sezione di Torino, Alessandria, Italy
\item \Idef{org33}Dipartimento Interateneo di Fisica `M.~Merlin' and Sezione INFN, Bari, Italy
\item \Idef{org34}Division of Experimental High Energy Physics, University of Lund, Lund, Sweden
\item \Idef{org35}European Organization for Nuclear Research (CERN), Geneva, Switzerland
\item \Idef{org36}Excellence Cluster Universe, Technische Universit\"{a}t M\"{u}nchen, Munich, Germany
\item \Idef{org37}Faculty of Engineering, Bergen University College, Bergen, Norway
\item \Idef{org38}Faculty of Mathematics, Physics and Informatics, Comenius University, Bratislava, Slovakia
\item \Idef{org39}Faculty of Nuclear Sciences and Physical Engineering, Czech Technical University in Prague, Prague, Czech Republic
\item \Idef{org40}Faculty of Science, P.J.~\v{S}af\'{a}rik University, Ko\v{s}ice, Slovakia
\item \Idef{org41}Faculty of Technology, Buskerud and Vestfold University College, Tonsberg, Norway
\item \Idef{org42}Frankfurt Institute for Advanced Studies, Johann Wolfgang Goethe-Universit\"{a}t Frankfurt, Frankfurt, Germany
\item \Idef{org43}Gangneung-Wonju National University, Gangneung, Republic of Korea
\item \Idef{org44}Gauhati University, Department of Physics, Guwahati, India
\item \Idef{org45}Helmholtz-Institut f\"{u}r Strahlen- und Kernphysik, Rheinische Friedrich-Wilhelms-Universit\"{a}t Bonn, Bonn, Germany
\item \Idef{org46}Helsinki Institute of Physics (HIP), Helsinki, Finland
\item \Idef{org47}Hiroshima University, Hiroshima, Japan
\item \Idef{org48}Indian Institute of Technology Bombay (IIT), Mumbai, India
\item \Idef{org49}Indian Institute of Technology Indore, Indore, India
\item \Idef{org50}Indonesian Institute of Sciences, Jakarta, Indonesia
\item \Idef{org51}INFN, Laboratori Nazionali di Frascati, Frascati, Italy
\item \Idef{org52}INFN, Sezione di Bari, Bari, Italy
\item \Idef{org53}INFN, Sezione di Bologna, Bologna, Italy
\item \Idef{org54}INFN, Sezione di Cagliari, Cagliari, Italy
\item \Idef{org55}INFN, Sezione di Catania, Catania, Italy
\item \Idef{org56}INFN, Sezione di Padova, Padova, Italy
\item \Idef{org57}INFN, Sezione di Roma, Rome, Italy
\item \Idef{org58}INFN, Sezione di Torino, Turin, Italy
\item \Idef{org59}INFN, Sezione di Trieste, Trieste, Italy
\item \Idef{org60}Inha University, Incheon, Republic of Korea
\item \Idef{org61}Institut de Physique Nucl\'eaire d'Orsay (IPNO), Universit\'e Paris-Sud, CNRS-IN2P3, Orsay, France
\item \Idef{org62}Institute for Nuclear Research, Academy of Sciences, Moscow, Russia
\item \Idef{org63}Institute for Subatomic Physics of Utrecht University, Utrecht, Netherlands
\item \Idef{org64}Institute for Theoretical and Experimental Physics, Moscow, Russia
\item \Idef{org65}Institute of Experimental Physics, Slovak Academy of Sciences, Ko\v{s}ice, Slovakia
\item \Idef{org66}Institute of Physics, Academy of Sciences of the Czech Republic, Prague, Czech Republic
\item \Idef{org67}Institute of Physics, Bhubaneswar, India
\item \Idef{org68}Institute of Space Science (ISS), Bucharest, Romania
\item \Idef{org69}Institut f\"{u}r Informatik, Johann Wolfgang Goethe-Universit\"{a}t Frankfurt, Frankfurt, Germany
\item \Idef{org70}Institut f\"{u}r Kernphysik, Johann Wolfgang Goethe-Universit\"{a}t Frankfurt, Frankfurt, Germany
\item \Idef{org71}Institut f\"{u}r Kernphysik, Westf\"{a}lische Wilhelms-Universit\"{a}t M\"{u}nster, M\"{u}nster, Germany
\item \Idef{org72}Instituto de Ciencias Nucleares, Universidad Nacional Aut\'{o}noma de M\'{e}xico, Mexico City, Mexico
\item \Idef{org73}Instituto de F\'{i}sica, Universidade Federal do Rio Grande do Sul (UFRGS), Porto Alegre, Brazil
\item \Idef{org74}Instituto de F\'{\i}sica, Universidad Nacional Aut\'{o}noma de M\'{e}xico, Mexico City, Mexico
\item \Idef{org75}IRFU, CEA, Universit\'{e} Paris-Saclay, Saclay, France
\item \Idef{org76}iThemba LABS, National Research Foundation, Somerset West, South Africa
\item \Idef{org77}Joint Institute for Nuclear Research (JINR), Dubna, Russia
\item \Idef{org78}Konkuk University, Seoul, Republic of Korea
\item \Idef{org79}Korea Institute of Science and Technology Information, Daejeon, Republic of Korea
\item \Idef{org80}KTO Karatay University, Konya, Turkey
\item \Idef{org81}Laboratoire de Physique Subatomique et de Cosmologie, Universit\'{e} Grenoble-Alpes, CNRS-IN2P3, Grenoble, France
\item \Idef{org82}Lawrence Berkeley National Laboratory, Berkeley, California, United States
\item \Idef{org83}Moscow Engineering Physics Institute, Moscow, Russia
\item \Idef{org84}Nagasaki Institute of Applied Science, Nagasaki, Japan
\item \Idef{org85}National and Kapodistrian University of Athens, Physics Department, Athens, Greece
\item \Idef{org86}National Centre for Nuclear Studies, Warsaw, Poland
\item \Idef{org87}National Institute for Physics and Nuclear Engineering, Bucharest, Romania
\item \Idef{org88}National Institute of Science Education and Research, HBNI, Jatni, India
\item \Idef{org89}National Nuclear Research Center, Baku, Azerbaijan
\item \Idef{org90}National Research Centre Kurchatov Institute, Moscow, Russia
\item \Idef{org91}Niels Bohr Institute, University of Copenhagen, Copenhagen, Denmark
\item \Idef{org92}Nikhef, Nationaal instituut voor subatomaire fysica, Amsterdam, Netherlands
\item \Idef{org93}Nuclear Physics Group, STFC Daresbury Laboratory, Daresbury, United Kingdom
\item \Idef{org94}Nuclear Physics Institute, Academy of Sciences of the Czech Republic, \v{R}e\v{z} u Prahy, Czech Republic
\item \Idef{org95}Oak Ridge National Laboratory, Oak Ridge, Tennessee, United States
\item \Idef{org96}Petersburg Nuclear Physics Institute, Gatchina, Russia
\item \Idef{org97}Physics Department, Creighton University, Omaha, Nebraska, United States
\item \Idef{org98}Physics department, Faculty of science, University of Zagreb, Zagreb, Croatia
\item \Idef{org99}Physics Department, Panjab University, Chandigarh, India
\item \Idef{org100}Physics Department, University of Cape Town, Cape Town, South Africa
\item \Idef{org101}Physics Department, University of Jammu, Jammu, India
\item \Idef{org102}Physics Department, University of Rajasthan, Jaipur, India
\item \Idef{org103}Physikalisches Institut, Eberhard Karls Universit\"{a}t T\"{u}bingen, T\"{u}bingen, Germany
\item \Idef{org104}Physikalisches Institut, Ruprecht-Karls-Universit\"{a}t Heidelberg, Heidelberg, Germany
\item \Idef{org105}Physik Department, Technische Universit\"{a}t M\"{u}nchen, Munich, Germany
\item \Idef{org106}Research Division and ExtreMe Matter Institute EMMI, GSI Helmholtzzentrum f\"ur Schwerionenforschung GmbH, Darmstadt, Germany
\item \Idef{org107}Rudjer Bo\v{s}kovi\'{c} Institute, Zagreb, Croatia
\item \Idef{org108}Russian Federal Nuclear Center (VNIIEF), Sarov, Russia
\item \Idef{org109}Saha Institute of Nuclear Physics, Kolkata, India
\item \Idef{org110}School of Physics and Astronomy, University of Birmingham, Birmingham, United Kingdom
\item \Idef{org111}Secci\'{o}n F\'{\i}sica, Departamento de Ciencias, Pontificia Universidad Cat\'{o}lica del Per\'{u}, Lima, Peru
\item \Idef{org112}SSC IHEP of NRC Kurchatov institute, Protvino, Russia
\item \Idef{org113}Stefan Meyer Institut f\"{u}r Subatomare Physik (SMI), Vienna, Austria
\item \Idef{org114}SUBATECH, IMT Atlantique, Universit\'{e} de Nantes, CNRS-IN2P3, Nantes, France
\item \Idef{org115}Suranaree University of Technology, Nakhon Ratchasima, Thailand
\item \Idef{org116}Technical University of Ko\v{s}ice, Ko\v{s}ice, Slovakia
\item \Idef{org117}Technical University of Split FESB, Split, Croatia
\item \Idef{org118}The Henryk Niewodniczanski Institute of Nuclear Physics, Polish Academy of Sciences, Cracow, Poland
\item \Idef{org119}The University of Texas at Austin, Physics Department, Austin, Texas, United States
\item \Idef{org120}Universidad Aut\'{o}noma de Sinaloa, Culiac\'{a}n, Mexico
\item \Idef{org121}Universidade de S\~{a}o Paulo (USP), S\~{a}o Paulo, Brazil
\item \Idef{org122}Universidade Estadual de Campinas (UNICAMP), Campinas, Brazil
\item \Idef{org123}Universidade Federal do ABC, Santo Andre, Brazil
\item \Idef{org124}University of Houston, Houston, Texas, United States
\item \Idef{org125}University of Jyv\"{a}skyl\"{a}, Jyv\"{a}skyl\"{a}, Finland
\item \Idef{org126}University of Liverpool, Liverpool, United Kingdom
\item \Idef{org127}University of Tennessee, Knoxville, Tennessee, United States
\item \Idef{org128}University of the Witwatersrand, Johannesburg, South Africa
\item \Idef{org129}University of Tokyo, Tokyo, Japan
\item \Idef{org130}University of Tsukuba, Tsukuba, Japan
\item \Idef{org131}Universit\'{e} Clermont Auvergne, CNRS/IN2P3, LPC, Clermont-Ferrand, France
\item \Idef{org132}Universit\'{e} de Lyon, Universit\'{e} Lyon 1, CNRS/IN2P3, IPN-Lyon, Villeurbanne, Lyon, France
\item \Idef{org133}Universit\'{e} de Strasbourg, CNRS, IPHC UMR 7178, F-67000 Strasbourg, France, Strasbourg, France
\item \Idef{org134}Universit\`{a} degli Studi di Pavia, Pavia, Italy
\item \Idef{org135}Universit\`{a} di Brescia, Brescia, Italy
\item \Idef{org136}V.~Fock Institute for Physics, St. Petersburg State University, St. Petersburg, Russia
\item \Idef{org137}Variable Energy Cyclotron Centre, Kolkata, India
\item \Idef{org138}Warsaw University of Technology, Warsaw, Poland
\item \Idef{org139}Wayne State University, Detroit, Michigan, United States
\item \Idef{org140}Wigner Research Centre for Physics, Hungarian Academy of Sciences, Budapest, Hungary
\item \Idef{org141}Yale University, New Haven, Connecticut, United States
\item \Idef{org142}Yonsei University, Seoul, Republic of Korea
\item \Idef{org143}Zentrum f\"{u}r Technologietransfer und Telekommunikation (ZTT), Fachhochschule Worms, Worms, Germany
\end{Authlist}
\endgroup
\end{document}